\documentclass[a4paper,11pt]{article}
\usepackage{stmaryrd}
\usepackage{amsthm,amssymb,amsmath}
\usepackage{algorithm}
\usepackage{algorithmic}
\theoremstyle{definition}
\newtheorem{definition}{Definition}[section]
\usepackage{hyperref}
\usepackage{svg}
\usepackage{ulem}
\usepackage{float}
\usepackage{graphicx}
\newcommand{\R}{\mathbb{R}}

\providecommand{\keywords}[1]
{
  \small	
  \textbf{\textit{Keywords---}} #1
}

\begin{document}
\title{Biometric Masterkeys}
\author{Tanguy~Gernot, and Patrick~Lacharme \\
 \small Normandie~Univ,~UNICAEN,~ENSICAEN,~CNRS,~GREYC,~14000~Caen,~France}

\maketitle

\begin{abstract}
Biometric authentication is used to secure digital or physical access. Such an 
authentication system uses a biometric database, where data are sometimes 
protected by cancelable transformations. This paper introduces the notion of 
biometric masterkeys. A masterkey is a feature vector such that the 
corresponding template matches with a significant number of templates 
stored in a cancelable biometric database. 
Such a masterkey is directly researched from a cancelable biometric database, but 
we also investigate 
another scenario in which the masterkey is fixed before the creation of the 
cancelable biometric database, providing additional access rights in the 
system for the masterkey's owner. 
Experimental results on the fingerprint 
database FVC and the face image database LFW show the effectiveness and 
the efficiency of such masterkeys in both scenarios. 
In particular, from any given feature vector, we are able to construct a 
cancelable database, for which the biometric template matches
with all the templates of the database. 
\end{abstract}

\section{Introduction}

Nowadays, biometrics is more and more used in many applications, from
access control on smartphones to payment on the Internet.
Biometrics is generally based on physical characteristics such 
as fingerprints, face, iris or voice. 
Several ways are used for biometric data protection, as cryptography or 
cancelable biometric transformations. These transformations 
should respect several security properties as revocability, unlinkability and 
non-invertibility while maintaining good performance. They use a secret
parameter, called token or seed, but the most used security model considers 
that this secret can be known by an attacker (stolen token model). 

During the enrollment step, one or several biometric data is transformed into 
a biometric template and stored in a database with biometric templates of 
other people. During the authentication step, a person sends a biometric data 
and the seed to the matcher, the corresponding template is computed and 
compared with templates of the database. If the matching is correct, the 
person is authenticated.
The performance of the cancelable scheme is expected close to the original 
scheme without transformations, which means similar biometric data are
mapped to similar templates with high probability. 

In this paper, we are interested in the creation of feature vectors for which 
the corresponding templates match with the greatest number of templates of the 
cancelable database. Such feature vectors, called masterkeys, can be used to 
attack the system, in the stolen token model, or provide some additional 
access rights for the system administrator.  
In this case, the cancelable transformation is used 
not for the security of biometric data, but for the creation of efficient 
masterkeys, by considering the secret seed as a public salt.
More precisely, masterkeys are computed in the two following scenarios:
\begin{itemize}
\item The cancelable database and the corresponding seeds are known. 
The goal is to create a synthetic feature vector for which the
corresponding templates, computed with the seeds, matches with the
maximal number of templates of the database.
\item The biometric database is known, and a feature vector is fixed.
The goal is to find the seeds for which the
corresponding templates, computed with the seeds, match with the
maximal number of templates, computed with the same seeds.
\end{itemize}
Experiments use a random projection, with or without a binarization step, for 
transformation, applied on two biometric databases: the FVC database 
(fingerprints) and the LFW database (face images).

In the second scenario, a brute force strategy is used for the computation of 
seeds, providing successful masterkeys for all the biometric database, with 
no significative performance degradation in these cancelable biometric 
databases. This second scenario proposes a new paradigm in biometric 
authentication, with the creation of biometric databases with specific access 
rights. It can also be seen as a backdoor placed in the system.
Seeds are also computed for a set of masterkeys's candidate, providing
partial masterkeys for other feature vectors.

Section \ref{sec:soa} proposes generic definitions on biometric systems and 
presents the works close to this paper. Section \ref{sec:theorical} formalizes 
the concept of masterkey with several definitions on the effectiveness of
such masterkeys.
Experiments and results are given in Section \ref{sec:results}, whereas
conclusion and future works are in Section \ref{sec:conclusion}.


\section{Definitions and related works}\label{sec:soa}

A biometric authentication scheme is realized into two phases. One (or several) 
feature vector is extracted from a biometric data during the enrollment process 
and is stored into a biometric database. During the authentication phase, a second 
feature vector is extracted from a biometric data and these two vectors are compared. 
 
\begin{definition}
Let $(\mathcal{M}_A, D_A)$ be a metric space. A biometric authentication scheme 
for two biometric data $b$ and $b'$ is a pair of algorithms $(E,V)$, where
\begin{itemize}
\item $E$ takes biometric data $b$ as input, and returns a feature vector 
$x \in \mathcal{M}_A$.
\item $V$ takes two feature vectors $x=E(b)$, ${x^\prime}=E(b')$, and a 
threshold $\tau_A$ as input, and returns $True$ if 
$D_A(x, {x^\prime})<\tau_A$, and $False$ otherwise.
\end{itemize}
\end{definition}

It is expected that feature vectors $x$ and $x^\prime$ of the same individual, 
extracted from two measurements are sufficiently close to verify 
$D_A(x, {x^\prime}) < \tau_A$. This paper uses the Euclidean distance for 
$D_A$, defined for feature vectors $x,x^\prime\in \mathcal{M}_A$ = ${\R}^N$ 
by: $$D_A(x,{x^\prime})=\sqrt{\sum_{i=1}^{N}{(x_i-{x^\prime}_i)^2}}.$$

\begin{definition}
Let $B=\{x_i\}_{i=1,\ldots ,n}$ be a biometric database, composed of feature 
vectors. A biometric authentication scheme for $B$ is an algorithm that 
takes a feature vector $x$ and a threshold $\tau_A$ as inputs and returns $True$ 
if $\exists i$ between $1$ and $n$ such that $V(x,x_i,\tau_A)$ returns $True$
and $False$ otherwise.
In the first case, $x$ is successfully authenticated by $x_i$.
\end{definition}

The quality of a biometric database is estimated with two rates: the False 
Match Rate (FMR) that provides the percentage of impostors authenticated by 
the system and the False Non Match Rate (FNMR) that provides the percentage 
of genuine peoples rejected by the system. These rates depend on $\tau_A$ 
and in case of equality these rates correspond to the Equal Error Rate (EER).

A biometric database contains some feature vectors that strongly influence 
these indicators. Doddington et al. \cite{doddington_sheep_nodate} sorted 
them in 4 categories: sheep, goats, lambs and wolves, where wolves are 
able to usurp many users, causing false acceptances. These individuals 
appear to have generic characteristics. 
Similarly, Yager et al. \cite{yager_biometric_2010} described 4 categories 
of animals representing the relationship between genuine and impostor match 
scores: worms, chameleons, phantoms and doves, where chameleons combine 
both high genuine and impostor match scores. In these categories, 
chameleons obtain scores belonging to the first quarter of the best scores.

Ine et al. defined a wolf as a feature vector that has a successful
authentication probability greater than the FMR and evaluate the security
of a given database with the notion of Wolf Attack Probability (WAP) in~\cite{ioi07}.
Countermeasures against this attack is investigated by Inuma et al.
in~\cite{ioi09} and tested on fingerprints by Murakami et al. 
in~\cite{mtm12}. Wolf attacks have been realized on the voice by Okhi et al.
in~\cite{oht12}, then by Ohki and Otsuka~\cite{oo14}.

Recently, Roy et al. investigated masterprints, partial fingerprints that usurp 
a large number of users~\cite{roy_masterprint_2017} in a multi-enrollment
process. They worked with small 
sensors of smartphones which acquire a lot of partial  
minutiae-based feature vectors. It was later improved with machine 
learning based approaches 
in~\cite{roy_evolutionary_2018, roy_masterprint_2019, bontrager_deepmasterprints_2018} 
and recently applied on face recognition in~\cite{nyem20}.

For security reasons, feature vectors are not always directly stored in the 
database, but are used as input of a cancelable biometric transformation.
These transformations were first proposed by Ratha et al. for face
recognition \cite{rcb01}. There are many works on cancelable transformations, as
presented in the surveys~\cite{ru11,nj15} on biometric templates protection.

\begin{definition}
Let $\mathcal{K}$ be a token (seed) space and $(\mathcal{M}_B, D_B)$ be a metric space.
A cancelable biometric scheme is a set of algorithms $(\mathcal{T}, \mathcal{V})$, 
where
\begin{itemize}
\item $\mathcal{T}$ takes a secret seed $s\in \mathcal{K}$, and a feature 
vector $x \in \mathcal{M}_A$ as input, and returns a biometric template 
$u=\mathcal{T}(s,x) \in \mathcal{M}_B$.
\item $\mathcal{V}$ takes two biometric templates $u$ = $\mathcal{T}(s,x)$, 
${u^\prime}=\mathcal{T}(s,{x^\prime})$, and a threshold $\tau_B$ as input, and 
returns $True$ if $D_B(u, {u^\prime}) < \tau_B$, and $False$ otherwise.
\end{itemize} 
\end{definition}

\begin{definition}\label{def:cbd}
Let $D$ = $\{u_i\}_{i=1,\ldots , n}$ be a cancelable biometric database, composed
of biometric templates $u_i$ = $\mathcal{T} (s_i,x_i)$. A cancelable biometric 
authentication scheme for $D$ is an algorithm that takes a biometric template 
$u$ = $\mathcal{T}(s,x)$ and a threshold $\tau_B$ as inputs and returns $True$ 
if $\exists i$ between $1$ and $n$ such that $s$ = $s_i$ and 
$\mathcal{V}(u,\mathcal{T}(s_i,x_i), \tau_B)$ returns $True$ and $False$ 
otherwise. In the first case, $x$ is successfully authenticated by $x_i$.
\end{definition}

It is expected that the cancelable biometric transformation $\mathcal{T}$ does 
not significantly decrease the performance of the original biometric system, 
in terms of EER. This paper uses the Hamming distance for $D_B$, defined for 
biometric templates $u,u^\prime\in \mathcal{M}_B$ = $\{0,1\}^M$  by: 
$$D_B(u,{u^\prime})=\sum_{i=1}^{M}{u_i\oplus u^\prime_i},$$ where $\oplus$ is 
exclusive or (xor) bitwise operator.\\

Cancelable biometric transformations are generally based on locality sensitive
hashing~\cite{a02,ai06,wzsss18} or robust perceptual hashing~\cite{
vkbp09, vkbp09bis}, ensuring the performance of the transformation. 
The transformations used in this paper are based on random projection, possibly 
with a binarization step. These cancelable transformations
were first proposed in \cite{teoh_personalised_2004,tkl08} with
the biohashing algorithm. Other projections have been later proposed, as
in~\cite{fyj10,ppcr11,wp10}. 

Let $M_s$ be a real matrix with $N$ rows and $M$ columns,
generated pseudo randomly from the random seed $s$, $x$ a feature vector of size $N$
and $D:{\R}^M\rightarrow \{0,1\}^M$ defined by $D(t_1,\ldots ,t_M)$ = 
$(u_1,\ldots ,u_M)$ where $$u_i= \left\{
\begin{array}{l}
0 \quad \mbox{if} \quad t_i< 0  \\
1 \quad \mbox{if} \quad t_i\geq 0 \\
\end{array}
\right.$$
In this case, the cancelable transformation $\mathcal{T}$ is defined by
$\mathcal{T}(x,s)$ = $D(xM_s)$.
In the biohashing algorithm, the pseudorandom matrix $M_s$ is orthonormalized
with the Gram-Schmidt algorithm.

The security of a cancelable biometric scheme is evaluated on several attack 
models, generally in the stolen token attack model, which considers that the 
token $s$ is known by the attacker.

\begin{definition} 
Let $x\in \mathcal{M}_A$ be a feature vector, 
$u=\mathcal{T}(s,x)\in\mathcal{M}_B$ is a biometric
template, computed from a seed $s$ and $\tau_A$ is a threshold. 
A nearby-feature preimage of $u$ with respect to $s$
is a feature vector $x^*$ such that $V(x,x^*,\tau_A)$ = $True$. 
\end{definition}

There are several methods for the construction of nearby-feature preimages. Genetic algorithms are 
a generic strategy for this, used in the case of biohashing on fingerprints in \cite{lacharme_preimage_nodate},
in the iris case in \cite{galbally_iris_2013}, in the case of minutiae templates in \cite{rozsa_genetic_2015}
or recently on several transformations in \cite{dong_genetic_2019}.
A genetic algorithm is a set of operators on a population, depending on a fitness function.
At each step, the population is updated with three operators: selection, crossing-over and 
mutation. Selection chooses the best subpart of the population. Crossing-over generates
new individuals in the population from the selected population. Mutation modifies the
new individuals generated during crossing-over. In the case of cancelable 
biometric transformations,
individuals are candidates for masterkey and the fitness function
is defined by the distance $D_B$ between the corresponding template and a target template.

Nearby-feature preimages can also be constructed using a specific attack
on the transformation. For example, random projection based schemes are vulnerable to
linear programming approaches~\cite{nnj10,fly14,tkae16}.


\section{Biometric masterkeys}
\label{sec:theorical}

A biometric masterkey $x$ is a synthetic feature vector such that the corresponding 
biometric template $u$ is close to several biometric templates in the 
database. The first part of this section is a generalization of concepts of
the previous section (wolves, masterprints) to cancelable biometric database.
The second part proposes an alternative by constructing a cancelable biometric
database that matches with a given masterkey. In this case, the cancelable 
transformation is introduced to provide such masterkeys efficiently. 

\begin{definition}
\label{def:masterkey}
Let $D = \{u_i\}_{i=1,\ldots ,n}$ be a cancelable biometric database 
and $\tau_B$ a threshold. $x$ is a masterkey for $D$, 
by respect to $\tau_B$, if $\forall i$ between $1$ and $n$,
$\mathcal{V}(\mathcal{T}(s_i,x),\mathcal{T}(s_i,x_i),\tau_B)$ = $True$, 
with $\mathcal{T}(s_i,x_i) = u_i$. 
\end{definition}

Obviously, $x$ is a masterkey for the database $\{\mathcal{T}(s,x)\}$, but 
the research of a masterkey for a large database is not necessarily an easy task. 
A less restrictive goal is the research of a minimal number of masterkeys 
such that all templates of the database are close to at less one
template corresponding to one masterkey.

\begin{definition}
\label{def:partition}
Let $D = \{u_i\}_{i=1,\ldots ,n}$ be a cancelable biometric database and 
$\tau_B$ a threshold. $D$ is said covered by a set of $r$ masterkeys 
$\{x^1,\ldots x^r\}$ by respect to $\tau_B$ if $\forall i$ between $1$ 
and $n$, $\exists k$ between $1$ and $r$ such that 
$\mathcal{V}(\mathcal{T}(s_i,x^k),\mathcal{T}(s_i,x_i),\tau_B)$ = $True$.\\
The minimum number $R$ such that $D$ is 
covered by a set of $R$ masterkeys is called optimal dictionary size.
\end{definition}

In the previous definition, each masterkey $x_i\in\{x_1,\ldots x_r\}$ is a 
masterkey for a subset $D_i\subset D$ and we have $\cup_{i=1}^rD_i=D$. 
Nevertheless, the number of templates in each subset $D_i$ is generally different. 

\begin{definition}
\label{def:covered}
A cancelable biometric database $D  = \{u_i\}_{i=1,\ldots ,n}$ is said 
$\epsilon$-covered by a masterkey $x$, with $0 < \epsilon\leq 1$ if 
there exists a subset $D'$ of $D$ of $\epsilon n$ persons such that $x$ is 
a masterkey for $D'$.
The optimal coverage percentage of $D$ is the maximal number 
$E$ such that $D$ is $E$-covered by a masterkey $x \in \mathcal{M}_A$. 
\end{definition}

Such a set of masterkeys can be used for attacks on the cancelable biometric 
database, similarly to the work of Roy et al. on 
minutiae~\cite{roy_masterprint_2017}, called dictionary attack, using
a set of carefully chosen masterkeys.
Nevertheless these masterkeys could provide other applications. For example,
a masterkey could be used for a system of delegation of biometric 
authentication, with the possibility for one person to be authenticated as 
herself and also as other person, similarly to proxy signatures~\cite{mso96}
or group signatures~\cite{cv91} in cryptography.\\
 
A second approach of the biometric masterkey problem is the construction of 
a cancelable biometric database from a feature vector $x$ and $n$ feature vectors 
extracted from $n$ peoples, such that $x$ is a masterkey for the cancelable
database. It means that, in this second approach, the seeds 
used by peoples are constructed, with the knowledge of $x$, contrary to the 
previous approach.  

\begin{definition}
\label{def:seed}
Let $x$ be a feature vector, $\tau_B$ be a threshold, and 
$B$ = $\{x_i\}_{i=1,\ldots ,n}$ be a biometric database.
A cancelable biometric database $D$ is said 
compliant with $x$ by respect to $\tau_B$ if there exists 
$s_1,\ldots s_n$ such that $D$ = $\{\mathcal{T}(s_i,x_i)\}$ and 
$x$ is a masterkey for $D$ by respect to $\tau_B$.
\end{definition}

This second approach uses not the cancelable biometric transformation for 
security but uses it for the construction of masterkeys. It makes possible to 
adjust the choice of seeds to allow super access, i.e., a vector usurping the 
entire base. This can allow an administrator to have the same access as its 
users without adding another authentication channel. If an insecure 
transformation is used, an additional security layer should be added in the 
template storage.

It can also allow an attacker active in the enrollment phase to build a 
backdoor in the authentication system,
without adding suspicious code or making the database dubious, as
in the following definition.

\begin{definition}
\label{def:backdoor}
Let $B$ = $\{x_i\}_{i=1,\ldots ,n}$ a biometric database and $D$ = 
$\{\mathcal{T}(s_i,x_i)\}$ a cancelable biometric database compliant
with a masterkey $x$. Then $x$ is a backdoor if there exists no
polynomial algorithm that distinguishes 
$s_1, \mathcal{T}(x_1, s_1), \ldots s_n, \mathcal{T}(x_n, s_n)$
from $s'_1, \mathcal{T}(x_1, s'_1), \ldots s'_n, \mathcal{T}(x_n, s'_n)$,
where $s'_1,\ldots s'_n$ are randomly generated from the set of seeds
($x$ is unknown).
\end{definition}

In this paragraph, we consider the theoretical complexity to find masterkeys.
We suppose that the metric space $(\mathcal{M}_B, D_B)$ is $\{0,1\}^M$ with
the Hamming distance and that the biometric transformation $\mathcal{T}$ is
pseudorandom. Thus, given a biometric template $u$, a feature vector $x$ and
a threshold $\tau$, the probability $p$ that the distance 
$D_B(\mathcal{T}(x,s), u) < \tau$
for a random seed $s$ is $p = \frac{1}{2^M}\sum_{i=0}^{\tau-1} \binom{M}{i}.$
Let $D  = \{u_i\}_{i=1,\ldots ,n}$ be a cancelable biometric database of
$n$ binary templates. A feature vector $x$, such that $D$ is $\epsilon$-covered 
by $x$, is expected after $1/p^{\epsilon n}$ trials.
Let $B = \{x_i\}_{i=1,\ldots ,n}$ be a biometric database of $n$ feature
vectors  and $x$ a feature vector.
A set of $n$ seeds $s_1,\ldots s_n$, such that the corresponding cancelable 
biometric database $D$ is compliant with $x$ (Definition~\ref{def:seed}) 
is expected after $n/p$ trials.
Thus, the research of a masterkey given a biometric database is clearly easier 
than given a cancelable database. 


\section{Experiments}\label{sec:results}

Effectiveness of masterkeys research is investigated on the fingerprint 
database FVC and the face database LFW. All experiments are processed on a 
Dell Latitude 7400 laptop with an Intel\textregistered~Core\texttrademark~i7-8665U 
CPU @ 1.90 GHz quad-core processor and 16 Gb of RAM, using python 
for the first part and C for the second one.


\subsection{Biometric Databases}

The first biometric database is the FVC2002 
database~\cite{fvc2002}, which contains $t$ = 8 biometric images of $n$ = 100 
peoples obtained from fingerprints. The second one is the sub LFW 
database~\cite{huang_labeled_2008} used in Dong et al. experiments 
\cite{dong_genetic_2019}, which contains $t$ = 10 biometric images of $n$ = 158 
peoples obtained from face pictures.
This database is noted LFW10 in this section. A third database, named LFW8, 
is extracted from LFW10 by taking the first 100 peoples, with the first 8 
biometric images, for comparison with the FVC2002 database.

Feature vectors are extracted from FVC using Gabor filters 
\cite{belguechi_comparative_2016}. Each biometric feature vector is 
composed of $N$ = $512$ real values of $64$ bits 
and the EER of the biometric database is $10\%$ with a threshold $\tau_A$ = $240.7$.
Feature vectors extracted from LFW are obtained using the deep network 
InsightFace \cite{deng_arcface_2019}. Each biometric feature vector is 
composed of $n$ = $512$ real values of $64$ bits and the EER of the biometric 
database is $0.2\%$ with a threshold $\tau_A$ = $1.227$.


\subsection{Masterkeys from Cancelable Biometric Database} 


The cancelable biometric transformation used in this subsection is the 
biohashing algorithm, mapping ${\R}^N$ into $\{0,1\}^M$ where $M$ = 128, 
described in Section~\ref{sec:soa}. The binary size 
of the seed is $128$ bits. After transformation the EER of the cancelable 
biometric database is around 16.5\% with a threshold $\tau_B$ = 17 
for FVC2002 and around 1.9\% with a threshold $\tau_B$ = 51
for LFW10 and LFW8.

The research of a masterkey, as presented in 
Definition~\ref{def:masterkey}, from a cancelable biometric database uses genetic 
algorithm where the fitness function is not the number of people who are 
authenticated by the masterkey, because it does not provide the best results. 
Two other fitness functions have been investigated, the first one based on 
the average of the minimal Hamming distance obtained for each person and the 
second one based on the sum of the minimal Hamming distance of not 
authenticated peoples by the masterkey. 
The second fitness function (\textit{sum}) provides the best results
on both databases and is used in next experiments.\\

We only consider the first biometric template for each person of
the cancelable biometric database during the evaluation of the best optimal 
coverage percentage and for the best optimal dictionary size.
 
\begin{figure}[H]
\centering
\includegraphics[width=9cm]{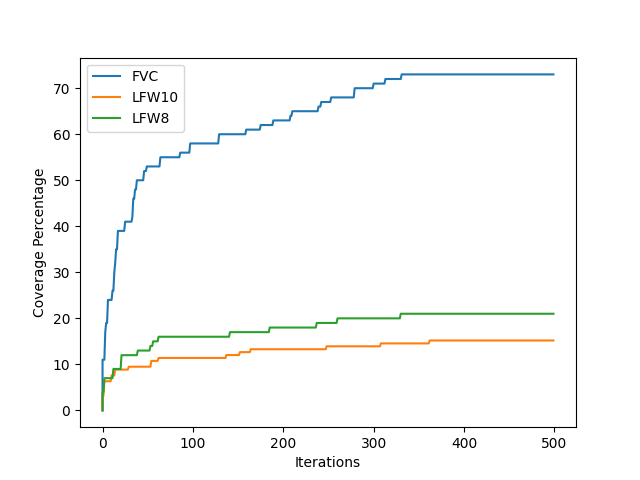}
\caption{Evolution of coverage during iterations}
\label{fig:time_iteration}
\end{figure}

Table \ref{tab:coverage} gives the best optimal coverage percentage (OCP),
as described in Definition~\ref{def:covered}, and the best optimal dictionary 
size (ODS), as described in Definition~\ref{def:partition}. For both 
cancelable biometric databases, we use thresholds $\tau_B$ at EER.
\begin{table}[H]
    \centering
\begin{tabular}{ |c|c|c| }
\hline
    db & OCP & ODS\\
\hline
   FVC  & 73 & 5\\
\hline
   LFW8 & 21 & 12\\
\hline
   LFW10 & 15.2 & 18\\
   \hline
\end{tabular}
    \caption{Optimal coverage percentage / Optimal dictionary size}
    \label{tab:coverage}
\end{table}
The genetic algorithm is processed on a population of $200$ peoples 
during $500$ iterations. Figure \ref{fig:time_iteration} shows that 
coverage does not increase from 350 iterations. 

The EER of the FVC database being important, lower thresholds for FVC
are presented in table \ref{tab:thresholdFAR} for a more accurate 
comparison of the OCP with LFW.
\begin{table}[H]
    \centering
\resizebox{\linewidth}{!}{
\begin{tabular}{ |c|c|c|c|c|c|c|c|c|c|c| }
\hline
Threshold $\tau_B$ & 16 & 15 & 14 & 13 & 12 & 11 & 10 & 9 & 8 & 7\\
\hline
FMR (\%) & 12.9 & 9.5 & 6.7 & 4.4 & 2.8 & 1.7 & 1 & 0.5 & 0.2 & 0.1\\
\hline
OCP & 64 & 58 & 49 & 42 & 38 & 29 & 23 & 19 & 14 & 11\\
\hline
\end{tabular}
}
\caption{FMR/OCP corresponding to the thresholds (FVC)}
\label{tab:thresholdFAR}
\end{table}

Vulnerabilities of cancelable transformations in the stolen token scenario have 
been already studied. Typically, nearby-feature preimages have 
been already constructed on biohashing in this scenario with various methods.
Nevertheless, it is the first time that these attacks are extended to the 
entire database by the construction of masterkeys. 
The strategy uses a genetic algorithm which is not related to the weaknesses
of this transformation. Typically, specific techniques as linear 
optimization algorithms give efficient results for nearby-feature preimages on 
biohashing and could be used to improve the performance of our algorithm. 
Nevertheless, the choice of this algorithm depends on the cancelable
transformation, whereas the research of masterkeys should be possible for
all transformations.


\subsection{Masterkeys from Biometric Database}  


In this second scenario, we have a biometric database and a feature vector,
called candidate masterkey. 
The objective is the generation of seeds for which the feature vector is a 
masterkey for the corresponding cancelable biometric database as presented in 
Definition~\ref{def:cbd}. In this scenario the transformation is chosen for
the efficiency of masterkeys creation, not for data protection. 
Thus, the beginning of this subsection describes the choice of the
transformation and its parameters.

The performance of random projection-based schemes are analyzed using the 
Johnson-Lindenstauss Lemma~\cite{jl84,dg03} which establishes that for any 
$0<\epsilon <1$, there exists a map $f:{\R}^N\rightarrow {\R}^M$ such that for 
all $x, y$ in a subset of $n$ points of ${\R}^N$ with $M\geq O(\epsilon -2\log n)$
$$(1-\epsilon )\parallel x - y\parallel_2^2\leq \parallel f(x) - f(y)\parallel_2^2
\leq (1+\epsilon )\parallel x - y\parallel_2^2.$$
The two following random projection $N\times M$ matrices are proposed by Achlioptas 
in \cite{a03}. In the first case, called JL1, coefficients of the projection 
matrix $M_s$ are 
$$\left\lbrace
\begin{array}{l}
1/\sqrt{M} \text{ with a probability 1/2}\\
-1/\sqrt{M} \text{ with a probability 1/2}
\end{array}\right.$$
In the second case, called JL2, coefficients of $M_s$ are 
$$\left\lbrace
\begin{array}{l}
\sqrt{3/M} \text{ with a probability 1/6}\\
0 \text{ with a probability 2/3}\\
-\sqrt{3/M} \text{ with a probability 1/6}
\end{array}\right.$$
The main advantage of these transformations is the low computational time 
for matrix generation compared to the Gram-Schmidt algorithm.

\begin{figure}[H]
\centering
\includegraphics[width=6cm]{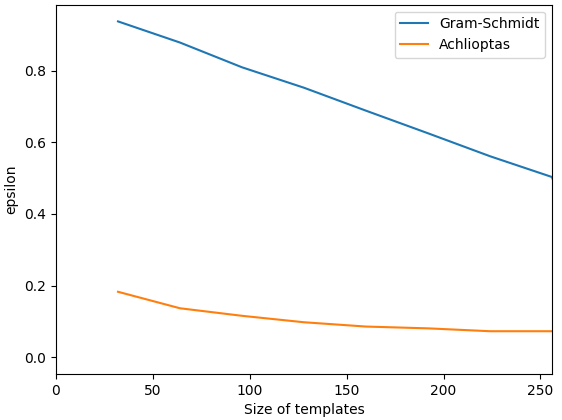}
\caption{Estimation of $\epsilon$ for different sizes of templates (FVC)}
\label{fig:epsilon}
\end{figure}

The value of $\epsilon$ in the previous inequality is detailed in 
figure~\ref{fig:epsilon} for $N$ = $512$ and $M$ between 32 and 256 using the 
FVC database (a random database also provides similar curves). Two 
transformations are used: the map composed of a random orthonormalized matrix 
(Gram-Schmidt), where $\epsilon $ decreases from $0.9$ to $0.5$ and for the 
two previous transformations (Achlioptas) which provides the similar results 
($\epsilon \simeq 0.1$ for $M\geq 128$).

\begin{figure}[H]
\centering
\includegraphics[width=6cm]{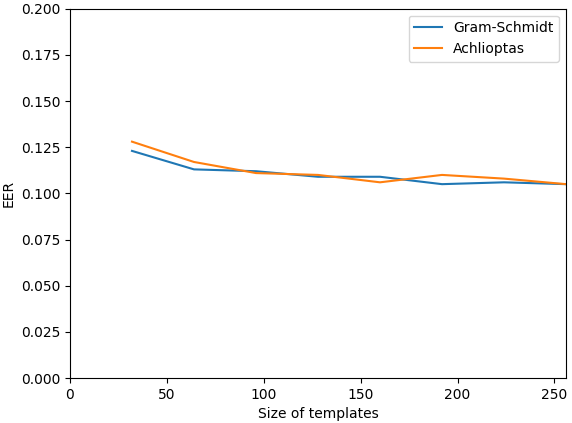}
\caption{EER for FVC (Euclidean distance)}
\label{fig:eer_euc}
\end{figure}

The previous experiments on $\epsilon$ has very few consequences on the EER.
The EER of the cancelable database is close to the EER of the original 
database in figure~\ref{fig:eer_euc} for both transformations for $M\geq 128$.
Even if the value of $\epsilon$ in the previous experiment is different,
in the case of Achlioptas's transformations, the threshold is stable, 
whereas in the first transformation, the threshold increases.
Thus, the research of masterkeys from a biometric database seems more 
interesting, directly in the Euclidean space, because the EER does not
change compared to the original database if $M$ is not too small
(for $M$ = $128$, around 11\% against 10\% on FVC and around 1\% against 
0.2\% for LFW). Nevertheless, the computational time for the research 
of masterkeys is not negligible, particularly for some feature vectors.
 
\begin{figure}[H]
\centering
\includegraphics[width=6cm]{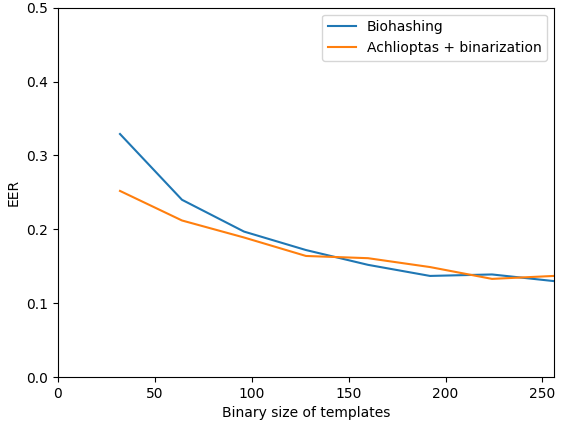}
\caption{EER for FVC (Hamming distance)}
\label{fig:eer_ham}
\end{figure}

The EER of the cancelable database obtained with the two previous 
transformations and a binarization step is a little bit 
increased compared to the original EER: around 16.5\% for FVC and 
2.1\% for LFW in figure~\ref{fig:eer_ham} with $M$ = $128$.
Consequently we investigate the research of masterkeys with transformations 
$\mathcal{T}:{\R}^N\rightarrow {\R}^M$ or $\{0,1\}^M$, defined by the Achlioptas 
transformations, ending or not by a binarization step. 
Random matrices are generated from a seed of $16$ bytes as previously.
 
Let $B=\{x_i^j\}_{i=1,\ldots ,n, j = 1\ldots ,t}$ be a biometric database, 
composed of feature vectors ($n$ peoples with $t$ vectors by people).
In experiments, the candidate masterkey is a feature vector of a chosen 
people of a database.
The research of seeds for a cancelable biometric database compliant with
a masterkey candidate $x_i^j$, as presented in Definition~\ref{def:seed}
is realized, taking one by one the $(n-1)$ $j^{th}$ feature vectors of each person, 
and choice seed, randomly generated from $\{0,1\}^{128}$,
 so that the candidate masterkey is successfully authenticated.
Experiments use the following algorithm, and outputs an array of $n-1$ seeds 
for each candidate masterkey, with the threshold at EER. 
The seed research is only limited by a value $c_{max}$, in the case of the
FVC database without binarization of the transformation.
If a successful seed is 
not founded after $c_{max}$ trials, the output of the algorithm is $Failure$
(in most of experiments, $c_{max}$ is not achieved).

\begin{center}
\begin{algorithm}[H]
\caption{Research of seeds for masterkeys} 
\begin{algorithmic}
\REQUIRE $B$ the database, $\tau_{B}$ the threshold, $c_{max}$\\
    \STATE SeedT = []
        \FOR{$i$ = $1$ to $n$ and $j$ = $1$ to $t$}
            \STATE SeedL = []
            \FOR{$k$ = 1 to $n$, with $k\neq i$}
                  \STATE $cpt\leftarrow 0$, $s_k\leftarrow rand(1, 2^{128})$
                   \WHILE{$\mathcal{V}(\mathcal{T}(s_k,x_i^j),\mathcal{T}(s_k,x_k^j),\tau_B)$ = $False$ and $cpt < c_{max}$} 
                       \STATE $s_k\leftarrow rand(1, 2^{128})$
                   \ENDWHILE
                   \IF{$\mathcal{V}(\mathcal{T}(s_k,x_i^j),\mathcal{T}(s_k,x_k^j),\tau_B)$ = $True$} 
                        \STATE SeedL.append($s_k$) 
                   \ELSE{} 
                         \STATE SeedL.append(Failure)
                   \ENDIF
            \ENDFOR
            \STATE SeedT.append(SeedL)
         \ENDFOR
    \STATE Return SeedT
\end{algorithmic}
\end{algorithm}
\end{center}

\begin{figure}
\includegraphics[width=7cm]{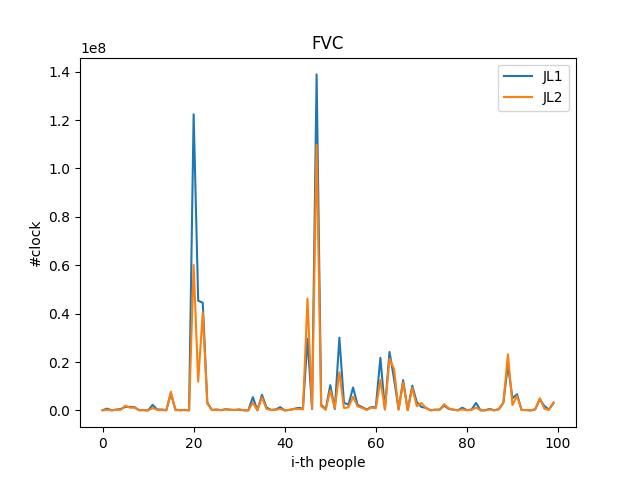}\hfill
\includegraphics[width=7cm]{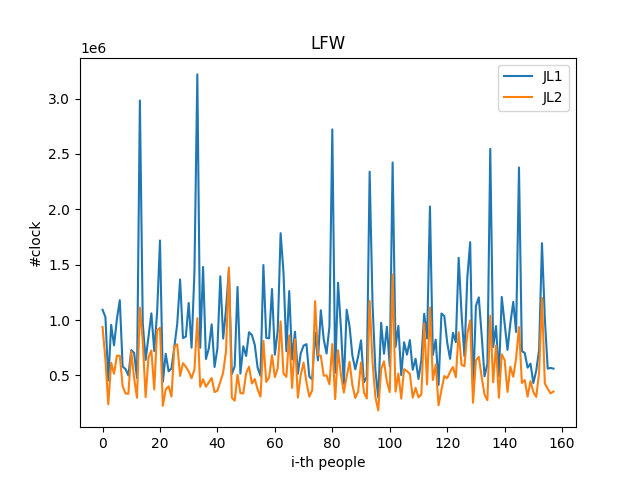}
\caption{Average number of clock (in binary)}
\label{fig:time_bin}
\end{figure}

The research of seeds for masterkeys is successful for all feature
vectors from FVC and LFW10 (called now LFW), for both Achlioptas's 
transformations (called JL1 and JL2), 
with the binarization step. Figure~\ref{fig:time_bin} presents the 
number of clock required for each
feature vector for both transformations on FVC database 
(more precisely, it is an average of the $t$ vectors of each person), 
and the same results for LFW.
It shows that the time is more important for some people of the 
FVC database, compared to the LFW database. It also shows that the
second transformation provides slightly better results in both database
(but the number of distance comparison is similar).
The research of seeds is very efficient and only takes a
few seconds for almost masterkey candidates.

The average of the EER in these $n*t$ cancelable biometric databases 
is around 17\% for FVC (for both transformations) which is 
close to the EER of a cancelable database obtained with random
seeds (around 16.5\% for both transformations). For LFW, the 
average of the EER is around 2.4 \% which is also close to the EER 
of a cancelable database obtained with random seeds
(around 2.1\% for both transformations). 

\begin{figure}[H]
\centering
\includegraphics[width=9cm]{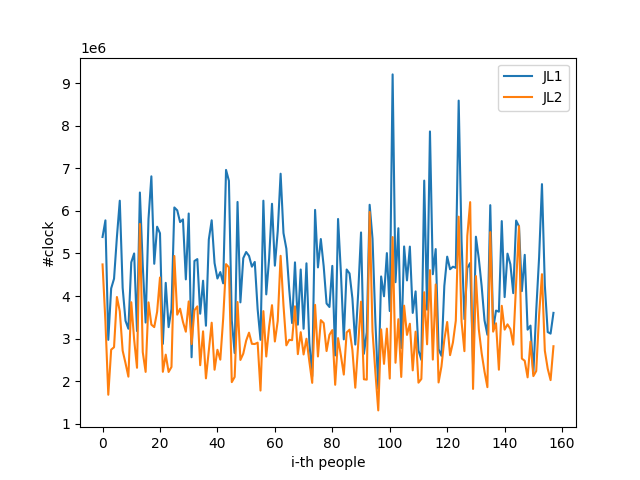}
\caption{Average number of clock (LFW in Euclidean)}
\label{fig:time_LFW_euc}
\end{figure}

If these transformations do not use a binarization step, the time
of research increases a lot for the FVC database (more precisely for
some people of this database), so we don't detail it. 
In the case of LFW database, the number of clock
is not significantly increased compared to the number of clock 
required with the binarization step.
Figure~\ref{fig:time_LFW_euc} presents the number of clock during the research of the masterkeys for LFW (more precisely, 
it is an average of the $t$ vectors of each person).
In this case, the second transformation is again more efficient than
the first one.
The EER of these $n*t$ databases is around 1.3\% which is close to the EER of 
the LFW database obtained with random seeds (around 1\% for both 
transformations without binarization). 

In this subsection, masterkeys have been efficiently founded for LFW
in the Euclidean space. Nevertheless, due to complexity for some vectors of 
the FVC database, it seems more appropriate to use a transformation
with binarization in the case where acquisition of
feature vectors has not been perfectly realized. There is a tradeoff between
performance of the cancelable database and efficiency of the research of
seeds for masterkeys.


\subsection{Multiple masterkeys}

In this subsection, a variant of the previous scenario is investigated.
The previous definition ensures for the owner's masterkey that the masterkey 
is successfully authenticated with enrollment data of other people. Nevertheless 
it does not ensure that another biometric data of the owner's masterkey is
authenticated on this database. This subsection studies the percentage 
of the cancelable database covered by these other feature vectors.\\ 

Let $B=\{x_i^j\}_{i=1,\ldots ,n, j = 1\ldots ,t}$ be a biometric database, 
composed of feature vectors ($n$ peoples with $t$ vectors by people). Let 
$x_1,\ldots x_t$ be feature vectors which are candidates for masterkey 
($x_1 ,\ldots x_t$ are assumed from the same people, but it doesn't mean 
that the distance between them is always lower than the used threshold).
In a first time seeds are computed for a subset $x_{t_1},\ldots x_{t_T}$,
called first set (if such a seed is not founded, the seed minimizing the worst distance is kept). In a second time, we look for the 
coverage percentage of the database $D$ provided by the $t-T$ other 
vectors, called second set. The corresponding templates of 
$x_1,\ldots , x_t$, noted $u_1,\ldots , u_t$, form the first set
and the second set of templates, depending to the feature vectors.

\begin{figure}[H]
\centering
\includegraphics[width=8.5cm]{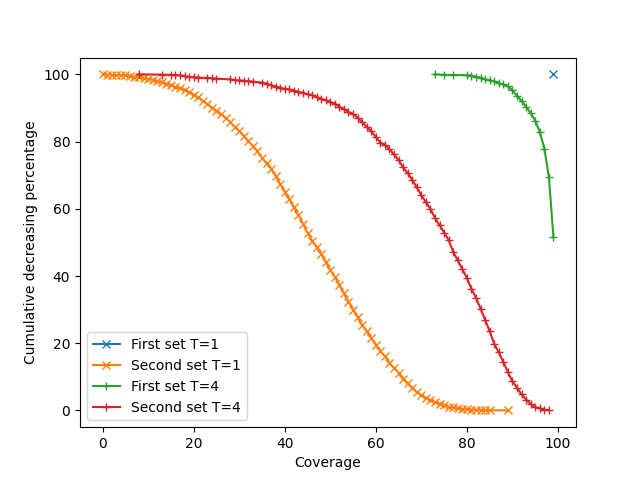}
\caption{Cumulative decreasing percentage of successful masterkeys for $T$ = $1$ and $4$,
on FVC (Binary)}
\label{fig:multiple_MK_binary}
\end{figure}

\begin{figure}[H]
\includegraphics[width=7cm]{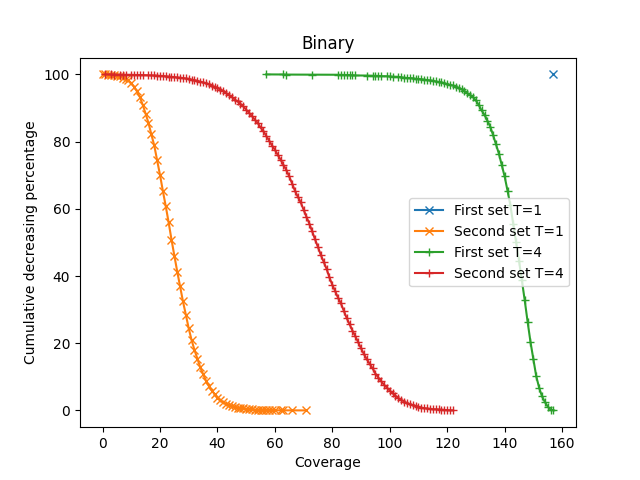}\hfill
\includegraphics[width=7cm]{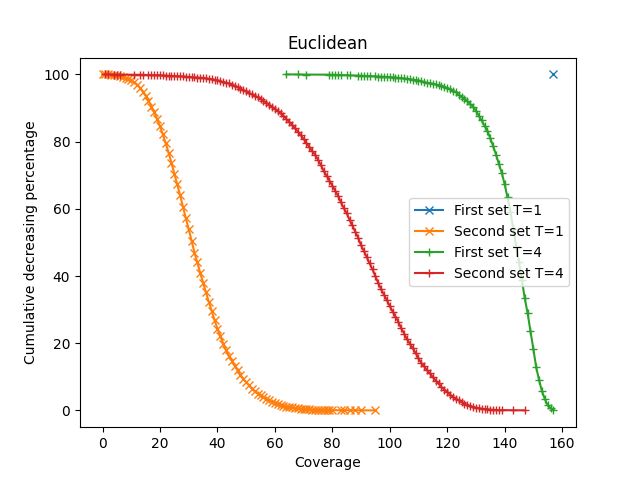}
\caption{Cumulative decreasing percentage of successful masterkeys for $T$ = $1$ and $4$,
on LFW (Binary and Euclidean)}
\label{fig:multiple_MK_euclidean}
\end{figure}

Experiments only use the JL2 transformation, with and without binarization,
on FVC and LFW, for $T$ = $1$ and $T$ = 4. For practical reasons, we used 
a time limit of 5 minutes in place of $c_{max}$. The number of templates in 
$D$ matching with the first set of $T$ templates 
$u_{t_1},\ldots u_{t_T}$ (in blue for $T$ = 1 and green for $T$ = 4) and 
with the second set of
$t- T$ other templates (in orange for $T$ = 1 and red for 
$T$ = 4) is presented in Figure \ref{fig:multiple_MK_binary}  
for FVC the Hamming space and in Figure \ref{fig:multiple_MK_euclidean} 
for LFW the Hamming and Euclidean space. 
 
For example, with $T=1$ in FVC, around $60\%$ of templates from the second 
set of templates match with at less $40$ templates of $D$ (orange curve), whereas 
with $T$ = 4, around $60\%$ of templates from the second set of templates 
match with at less $75$ templates of $D$ (red curve).
Moreover we have only one blue point 
in these figures, because 100\% of templates of the first set 
match with all templates in the database, as seen in the 
previous subsection (if $T$ = $4$ it is not the case because
the limit of 5 minutes for each seed computation).

Curves corresponding to the second set of templates show that, on 
average, seeds computed for 4 masterkey's candidate of the first set of 
feature vectors are better for the second set of templates than seeds 
computed for only one candidate of the first set of feature vectors. 

\begin{figure}[H]
\centering
\includegraphics[width=8.5cm]{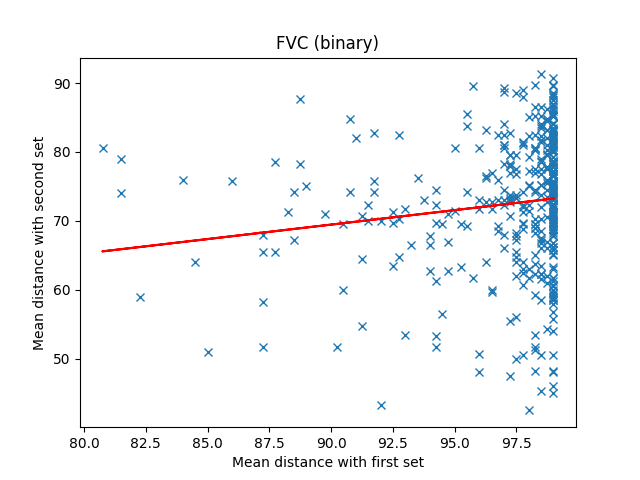}
\caption{Correlation between successful masterkeys and other templates
(T = 4 on FVC in Hamming space)}
\label{fig:correlation_fvc}
\end{figure}

\begin{figure}[H]
\includegraphics[width=7cm]{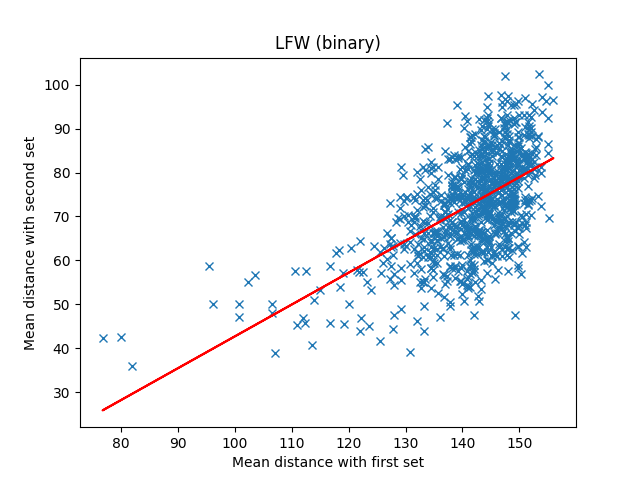}\hfill
\includegraphics[width=7cm]{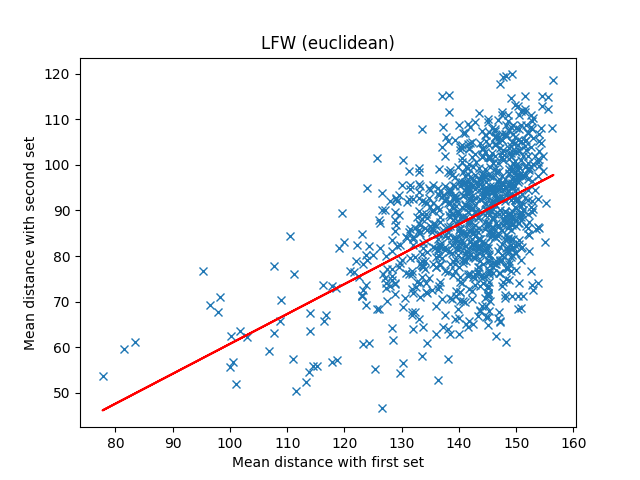}
\caption{Correlation between successful masterkeys and other templates
(T = 4 on LFW with Euclidean and Hamming space)}
\label{fig:correlation_lfw}
\end{figure}

Figures \ref{fig:correlation_fvc} and 
\ref{fig:correlation_lfw} present the correlation between the
number of successful masterkeys from the first set of feature vectors
(abscissa) and the number of successful 
matches of the other templates from the second set of feature vectors 
(ordinate). Red curves correspond to a linear regression of points. 
If the correlation is low in the case of FVC, it is really high in the
case of LFW, which confirms that the seeds should be computed from 
several acquisitions of feature vectors. Nevertheless, with $T$ =$4$,
the feature vectors of the first set are only masterkeys for a (large)
subset of the cancelable database, which provides a new tradeoff on
the value of $T$.

\section{Conclusion and Future Work}
\label{sec:conclusion}

This paper has introduced the concept of biometric masterkeys in two
scenarios. In the first one, there exists a cancelable biometric database,
and we try to find a feature vector for which the corresponding template
matches with the greatest number of templates of the database. It is a 
formalization and a generalization of Wolf Attacks and masterprints on 
fingerprints, experimented with the biohashing algorithm.
In the second scenario, there exists a biometric database,
and we try to find a set of seeds for which the corresponding template
of a given feature vector matches  with the greatest number of templates 
of the database.

The main contribution of this work is the construction of such a
cancelable biometric database covered by one of its feature vectors, 
with an EER similar to the original cancelable database.
The research of masterkeys in this second scenario is completely
different from previous works on wolves or masterprints, because it
uses the transformation, not for data protection, but for
 the research of preimages.
This work presents new opportunities for a biometric system 
administrator, to construct a cancelable biometric database which 
matches with a template obtained from its own feature vector.
This masterkey only requires a public seed by template and is 
very easy to construct.

There are several future works and improvements of this paper,
particularly in the research of seeds, providing good coverage
percentage of the cancelable database for other acquisitions 
of feature vectors. Other applications of masterkeys, as biometric 
systems with fine-grained access controls, will also be investigated 
in the future.
Finally, it would be interesting to use an efficient and secure
transformation to ensure the security of templates in this context.
Nevertheless, if such a transformation is not available, it is
possible to combine masterkeys with any biometric protection scheme.
\newpage
\bibliographystyle{splncs04}
\bibliography{main}  
\end{document}